\documentclass[twocolumn,showpacs,preprintnumbers,amsmath,amssymb,floatfix]{revtex4}

\usepackage{graphicx}
\usepackage{dcolumn}
\usepackage{bm}

\usepackage{amsmath}
\usepackage{psfrag}
\usepackage{color}
\usepackage{latexsym,amsfonts}
\usepackage{graphicx}
\providecommand{\href}[2]{#2}   
\definecolor{Blue2}{rgb}{0.,0.,0.8125}
\definecolor{Brown3}{rgb}{0.625,0.25,0.}
\definecolor{Cyan4}{rgb}{0.,0.56,0.56}
\definecolor{Green4}{rgb}{0.,0.56,0.}
\definecolor{LtBlue}{rgb}{0.27,0.42,0.52}
\definecolor{Magenta4}{rgb}{0.5625,0.,0.5625}
\definecolor{Red2}{rgb}{0.8125,0.,0.}

\begin{document}


\title{New Family of Exotic $\Theta$-Baryons}
\author{Dmitry Borisyuk$^1$}
 \email{borisyuk@ap3.bitp.kiev.ua}
\author{Manfried Faber$^2$}
 \email{faber@kph.tuwien.ac.at}
 
\author{Alexander~Kobushkin$^{1,2}$}%
 \email{akob@ap3.bitp.kiev.ua}
\affiliation{$^1$Bogolyubov Institute for Theoretical Physics, 03143, Kiev, Ukraine}
\affiliation{%
$^2$Atominstitut der \"Osterreichischen Universit\"aten,
        Technische Universit\"at Wien\\
         Wiedner Hauptstr. 8-10, A--1040 Vienna, Austria
}%

\date{\today}

\begin{abstract}
From the interpretation of the  $\Theta^+$ baryon resonance as an excitation of the ``skyrmion liquid'' with SU(3) flavor symmetry $\overline{10}$ we deduce a new series of baryons, $\Theta_1^{++}$, $\Theta_1^{+}$ and $\Theta_1^0$, situated at the top of the 27-plet of SU(3) flavor, with hypercharge $Y=2$, isospin $T=1$ and spin $J=\frac32$. The mass of $\Theta_1$ is estimated 55~$\mathrm{MeV/c^2}$ higher then the mass of $\Theta^+$ and its width at 80 MeV. We also discuss the other baryons from the 27-plet.  
\end{abstract}

\maketitle

Recently an exotic and narrow baryon resonance, $\Theta^+$, which cannot be formed by three quarks was observed in three independent experiments \cite{Nakano,Barmin,Stepanyan}. Masses of $1540\pm 10~\mathrm{MeV/c^2}$ \cite{Nakano}, $1539\pm 2~\mathrm{MeV/c^2}$ \cite{Barmin} and $1543\pm 5~\mathrm{MeV/c^2}$ \cite{Stepanyan} were reported, in excellent agreement with the theoretical prediction $M_{\Theta^+}^\mathrm{th}=1530~\mathrm{MeV/c^2}$ \cite{DPP}. In these experiments the resonance width was estimated at $<~25~\mathrm{MeV}$ \cite{Nakano}, $<~9~\mathrm{MeV}$ \cite{Barmin} and $<~22~\mathrm{MeV}$  \cite{Stepanyan}  comparable with the theoretical prediction $\Gamma_{\Theta^+}^\mathrm{th}< 15~\mathrm{MeV}$ \cite{DPP}. The theoretical predictions for the $\Theta^+$-baryon were done in the framework of the extended Skyrme model for the SU(3) flavor multiplet $\mu=(0,3)$ with the dimension $N_\mu=\overline{10}$ (anti-decuplet) \footnote{Some rough estimates for the  baryons from the anti-decuplet were also done in \cite{DP,Chemtob, BDothan,Praszalowicz,Walliser,Weigel}.}. The hypercharge of the observed $\Theta^+$, $Y=2$, follows from strangeness conservation in electromagnetic and strong interactions, the isospin cannot be determined from experiments \cite{Nakano,Barmin,Stepanyan}. If $\Theta^+$ is associated with the top of the anti-decuplet its other quantum numbers must be $T=0$ and $J^P=\frac12^+$.  

Contrary to the picture, where  $\Theta^+$ is considered as an excitation of a ``skyrmion liquid''  with appropriate SU(3) flavor symmetry, this resonance can also  be interpreted as Fock-state component $uudd\overline{s}$ \cite{CPR}. In this pure multi-quark picture $\Theta^+$ has isospin, spin and parity different from that predicted by the Skyrme model, e.g., $\Theta^+$ can be an isotensor resonance with $J^P=\frac12^-$, $\frac32^-$ or $\frac52^-$.

Besides exotic baryons in the anti-decuplet the Skyrme model predicts other SU(3) flavor multiplets with exotic baryons. A first estimate for the nearest partners of $\Theta^+$ shows that these must be exotic states in the $\mu=(2,2)$ representation (dimension $N_\mu=27$) with quantum numbers $Y=2$, $T=1$ and  $J^P=\frac32^+$. Depending on the fit of the known baryon spectra its mass was estimated between 100 and 150~$\mathrm{MeV/c^2}$ larger than the mass of the $\Theta^+$ \cite{WKop}. To clarify the situation it is important to make a detailed study of the predictions of the Skyrme model for the baryons from the anti-decuplet together with baryons from higher multiplets and to give, if possible, new predictions, which can support or reject the soliton picture for the nature of the $\Theta$-baryon. 

In the present short paper we calculate the mass spectrum of the baryons from the 27-plets with $J^P=\frac12^+$ and $\frac32^+$ and compare them with experiment. We find that the $J^P=\frac12^+$ baryons are systematically 500~$\mathrm{MeV/c^2}$ heavier than the $J^P=\frac32^+$ baryons. We show that besides two additional exotic resonances (which we call $\Gamma$ and $\Pi$) with hypercharge-isospin $(Y,T)=$~(0,2) and (-1,$\frac32$), respectively, there are new families of $\Theta$-baryons, $\Theta_1$ and $\Theta_2$, with $(Y,T)=$(2,1). The lightest of them should be only 55 $\mathrm{MeV/c^2}$ heavier than the $\Theta^+$-baryon. $\Theta_1$ has a typical hadronic width $\Gamma_{\Theta_1}\sim$80~MeV. 

Starting from a hedgehog ansatz and assuming rigid rotation in SU(3) space \cite{Witten,Guadagnini} one obtains the following Hamiltonian for the baryon representation $\mu=(p,q)$ of the SU(3) flavor group
\begin{equation}\label{Hamiltonian}
\begin{split}
H=&M_0 + \frac1{6 I_2}[p^2 + q^2 + pq + 3(p+q)] +\\
& +\left(\frac1{2I_1} - \frac1{2I_2}\right){\hat {\vec J}}^2 - \frac{(N_cB)^2}{24I_2} + \Delta \hat{H},
\end{split}
\end{equation}
where $\hat {\vec J}$ is the spin operator, $M_0$ is the energy of a static soliton solution, $I_1$ and $I_2$ are the two moments of inertia, $N_c=3$ is the number of colors and $B=1$ is the baryon number. All quantities $M_0$, $I_1$ and $I_2$ are functionals of the soliton profile. The Hamiltonian $\Delta \hat H$ is responsible for the splitting within SU(3) multiplets \cite{Guadagnini}
\begin{equation}\label{delta_H}
\begin{split}
\Delta \hat{H}(R) = \alpha D^{(8)}_{88}(R) + \beta Y + \frac\gamma{\sqrt3}\sum_{A=1}^3D^{(8)}_{8A}(R)\hat J_A.
\end{split}
\end{equation}
Here $D^{(8)}_{mn}(R)=\frac12 \mathrm{Tr}\left(R^\dag \lambda_m R \lambda_n\right)$ are Wigner rotation matrices for the adjoint SU(3) representation. The constants $\alpha$, $\beta$ and $\gamma$ are related to the current quark masses, $m_u$, $m_d$, $m_s$, the nucleon sigma term and four soliton moments of inertia \cite{DPP,Guadagnini}. 
\begin{figure}
  \psfrag{T3}{$T_3$}
  \psfrag{Y}{$Y$}
  \psfrag{Z1}{$\Theta_1$}
  \psfrag{D}{$\Delta,\ N$}
  \psfrag{S}{$\Gamma,\ \Sigma,\ \Lambda$}
  \psfrag{X}{$\Pi,\ \Xi$}
  \psfrag{O}{$\Omega$}
  \centering
  \includegraphics[height=0.3\textheight]{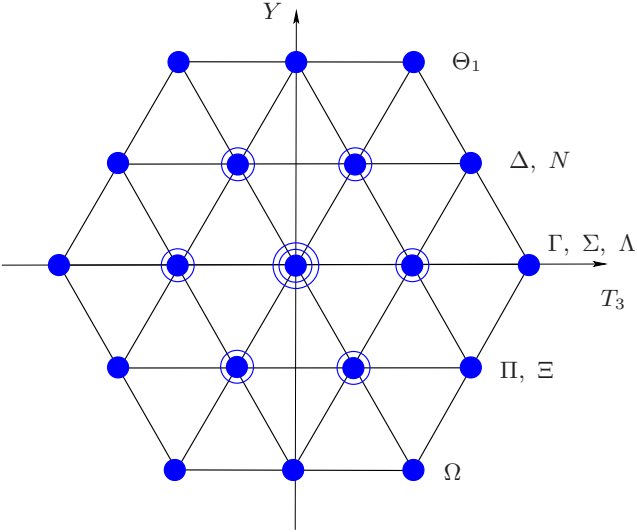}
  \caption{Structure of the 27-plet of baryons in the $T_3Y$ diagram.}
  \label{27-plet}
\end{figure}

Due to the Wess-Zumino term the quantization rule selects only such soliton spins $J$, which coincide with one of the allowed isospins $T$ for hypercharge $Y=1$ in the given SU(3) flavor multiplet \cite{Witten,Guadagnini}
\begin{equation}\label{selection}
J=T \qquad \text{for} \qquad Y=1.
\end{equation}  
So the lightest irreducible SU(3) representations which can be (very roughly) associated with 3-quark and 4-quark--antiquark systems and with allowed spins $J$ are
\begin{equation}\label{SU(3)-rep}
\begin{split}
\begin{array}{lll}
\text{octet}&\mu=(1,1)&J=1/2\\
\text{decuplet}&\mu=(3,0)&J=3/2\\
\text{anti-decuplet}&\mu=(0,3)&J=1/2\\
\text{27-plet}&\mu=(2,2)&J=1/2 \ \text{or} \ 3/2\\
\text{35-plet}&\mu=(4,1)&J=3/2 \ \text{or} \ 5/2
\end{array}
\end{split}
\end{equation}
The wave functions for baryons with hypercharge $Y$, isospin $T$, isospin 3-projection $T_3$, spin $J$ and its $z$-projection $J_3$ depend on 8 parameters (similar to Euler angles in SU(2)) of the SU(3) manifold
\begin{equation}\label{WFunc}
\langle R |\mu YTT_3JJ_3\rangle = \sqrt{N_\mu}(-1)^{J_3-\frac12}D^{\mu}_{YTT_3;1J-J_3}(R),
\end{equation}  
where $N_\mu$ is the dimension of the representation $\mu$. From the Hamiltonian (\ref{Hamiltonian}) we get the mass spectrum
\begin{equation}\label{sectrum}
\begin{split}
M=&M_0 + \frac1{6 I_2}[p^2 + q^2 + pq + 3(p+q)] +\\
& +\left(\frac1{2I_1} - \frac1{2I_2}\right)J(J+1) - \frac{3}{8I_2} +  \Delta M,
\end{split}
\end{equation}
where
\begin{eqnarray}\label{mass_split}
\Delta M&=&\langle \Delta\hat H \rangle = \\
&=&\int dR \langle\mu YTT_3JJ_3| R \rangle \Delta \hat{H}(R)  \langle R |\mu YTT_3JJ_3\rangle. \nonumber
\end{eqnarray}  
The rotational energy is given by the second and third terms in (\ref{sectrum}). In general it increases very strongly from the octet representation in (\ref{SU(3)-rep}) to the 35 representation. But there is one exception. From numerical results it follows that $I_1>I_2$. This means that in (\ref{sectrum}) the term proportional to $J(J+1)$ become more negative for higher angular momenta. So moving from the $J=\frac12$ anti-decuplet to the $J=\frac32$ 27-plet the increase of the rotational energy of the second term in (\ref{sectrum}) can be compensated by the increase of the negative contribution of the third term. Estimates with typical parameters for the moments of inertia $I_1$ and $I_2$ show that the rotation energy increases by $\approx 100$~MeV only, which, in principal, is of the order of the splitting within the SU(3) multiplet! The structure of the 27-plet of baryons in the $T_3Y$ diagram is displayed in Fig.~\ref{27-plet}. The states $\Theta_1$, $\Gamma_{27}$ and $\Pi_{27}$ are exotic and due to their $Y$ and/or $T$ values cannot be reduced to three quark systems. 

The SU(3) Clebsch-Gordan coefficients needed for calculations of the mass splitting in the 27-plet were taken from \cite{McNC}. These splittings are given in Table~\ref{tab:1} together with the results of Ref.~\cite{DPP} for the anti-decuplet. 

\begin{table}
\caption{\label{tab:1}Mass splitting in anti-decuplet with $J=\frac12$ and in 27-plets with $J=\frac12$ and $\frac32$.}
\begin{ruledtabular}
\begin{tabular}{cccc}
anti-decuplet\footnotemark[1]& $T$ & $Y$ & $\Delta M$\\
\hline
$\Theta^+$              & 0   & 2 & $ (1/4) \alpha +2\beta -(1/8) \gamma$\\
$N_{\overline{10}}$      & 1/2 & 1 & $ (1/8) \alpha + \beta -(1/16) \gamma$\\
$\Sigma_{\overline{10}}$ & 1   & 0 &                                    0\\ 
$\Xi_{\overline{10}}$    & 3/2 &-1 & $-(1/8) \alpha - \beta +(1/16) \gamma$\\
\hline
27-plet $J=3/2$\\
\hline
$\Theta_1$         & 1   & 2 & $ (1/7)    \alpha +2\beta -(5/14)   \gamma$\\
$\Delta_{27}$ & 3/2 & 1 & $ (13/112) \alpha + \beta -(65/224) \gamma$\\
$N_{27}$      & 1/2 & 1 & $ (1/28)   \alpha + \beta -(5/56)   \gamma$\\
$\Gamma_{27}$ & 2   & 0 & $ (5/56)   \alpha         -(25/112) \gamma$\\
$\Sigma_{27}$ & 1   & 0 & $-(1/56)   \alpha         +(5/112)  \gamma$\\
$\Lambda_{27}$& 0   & 0 & $-(1/14)   \alpha         +(5/28)   \gamma$\\
$\Pi_{27}$    & 3/2 &-1 & $-(1/14)   \alpha -  \beta +(5/28)  \gamma$\\
$\Xi_{27}$    & 1/2 &-1 & $-(17/112) \alpha -  \beta +(85/224)\gamma$\\
$\Omega_{27}$ & 1   &-2 & $-(13/56)  \alpha -2\beta +(65/112) \gamma$\\
\hline
27-plet $J=1/2$\\
\hline
$\Theta_1$         & 1   & 2 & $ (17/56)  \alpha +2\beta -(1/112)  \gamma$\\
$\Delta_{27}$ & 3/2 & 1 & $ (1/28)   \alpha + \beta -(5/112)  \gamma$\\
$N_{27}$      & 1/2 & 1 & $ (137/560)\alpha + \beta +(71/1120)\gamma$\\
$\Gamma_{27}$ & 2   & 0 & $-(13/56)  \alpha         -(19/112) \gamma$\\
$\Sigma_{27}$ & 1   & 0 & $ (13/280) \alpha         +(19/560) \gamma$\\
$\Lambda_{27}$& 0   & 0 & $ (13/70)  \alpha         +(19/140) \gamma$\\
$\Pi_{27}$    & 3/2 &-1 & $-(17/112) \alpha - \beta +(1/224)  \gamma$\\
$\Xi_{27}$    & 1/2 &-1 & $-(23/280) \alpha - \beta +(63/560) \gamma$\\
$\Omega_{27}$ & 1   &-2 & $ (1/14)   \alpha -2\beta +(5/28)   \gamma$
\end{tabular}
\end{ruledtabular}
\footnotetext[1]{From Ref.~\onlinecite{DPP}.}
\end{table}

In our numerical calculations we use the following parameters from Ref.~\cite{DPP}
\begin{equation}\label{Parameters}
\begin{split}
I_2&=(500~\mathrm{MeV})^{-1},\\
\alpha&=-218~\mathrm{MeV},\
\beta=-156~\mathrm{MeV},\
\gamma=-107~\mathrm{MeV}.
\end{split}
\end{equation}
The first moment of inertia $I_1$ was estimated from the experimental masses of the baryons from the $\frac12^+$ octet and the $\Sigma^\ast$ from the $\frac32^+$ decuplet
\begin{equation}\label{First_MOI} 
I_1=\frac2{3(m_{\Sigma^\ast}-m_\Sigma-m_N+m_\Lambda)}.
\end{equation}
$M_0$ we get from the mass of the nucleon. The estimated masses are given in Table~\ref{tab:2}. Concerning the $J^P=\frac12^+$ 27-plet we would like to mention that its states are approximately 500~$\mathrm{MeV}$ higher than the states of the $J^P=\frac32^+$ 27-plet.

Neglecting transitions to the 35-plet (which is around 1~GeV higher than the $J^P=\frac32^+$ 27-plet) the states 
\begin{equation}\label{Pure_states}
\Theta_1,\ N_{27},\ \Gamma_{27},\ \Lambda_{27},\ \Pi_{27},\ \text{and}\ \Omega_{27}\quad \text{with $J=\frac32$}
\end{equation}
exist as pure members of the 27-plet. The states $\Delta_{27}$, $\Sigma_{27}$ and $\Xi_{27}$ should be mixed with the corresponding decuplet states. Therefore, their wave functions read
\begin{equation}\label{Mixtire_states}
\begin{split}
|\Delta\rangle&=|\Delta_{10}\rangle + C_\Delta|\Delta_{27}\rangle,\\
|\Sigma\rangle&=|\Sigma_{10}\rangle + C_\Sigma|\Sigma_{27}\rangle,\\
|\Omega\rangle&=|\Omega_{10}\rangle + C_\Omega|\Omega_{27}\rangle,
\end{split}
\end{equation}
where the admixture coefficients are given by
\begin{equation}\label{admixtire_coef}
C_B=\frac{\langle B_{10}|\Delta\hat H |B_{27} \rangle}{M_{27}-M_{10}}, \quad M_{27}-M_{10}=\frac{1}{I_2}.
\end{equation}
The transition amplitudes read
\begin{equation}\label{Transition}
\begin{split}
\langle \Delta_{10}|\Delta\hat H |\Delta_{27} \rangle &=
\frac{\sqrt{30}}{16}\left(\alpha + \frac56 \gamma\right),\\
\langle \Sigma_{10}|\Delta\hat H |\Sigma_{27} \rangle &=
\frac{1}{4}\left(\alpha + \frac56 \gamma\right),\\
\langle\Xi_{10}|\Delta\hat H |\Xi_{27} \rangle &=
\frac{\sqrt{6}}{16}\left(\alpha + \frac56 \gamma\right).
\end{split}
\end{equation}
Using the parameters (8) and (9) one gets the following admixtures between the $J=\frac32$ 27-plet and the decuplet 
\begin{equation}\label{Admixture_numerically}
C_\Delta=-0.210,\ C_\Sigma=-0.154,\ C_\Xi=-0.094.
\end{equation}
This mixture for $J=\frac32$ baryons is huge, larger than the mixture between $J=\frac12$ baryons in the octet and the anti-decuplet which was shown to be universal and equal to $C_{8-10}=0.084$ \cite{DPP}. Such a strong mixture shows that one cannot ignore the $\overline{q}qqqq$ component in strong and electromagnetic transitions between nucleons and deltas.
\begin{table}
\caption{\label{tab:2}Mass spectrum in anti-decuplet with $J^P=\frac12^+$ and in 27-plet with  $J^P=\frac12^+$ and $J^P=\frac32^+$.}
\begin{ruledtabular}
\begin{tabular}{cccc}
particle          & $J$ &Theory ($\mathrm{MeV/c^2}$)    &Experiment\\
\hline
$\Theta^+$             & 1/2 & 1530 \footnotemark[5]                         &$\begin{cases}1540 \pm 10\footnotemark[1]\\1539 \pm 1\footnotemark[2] \\1542 \pm 5\footnotemark[3]\end{cases}$      \\
$N_{\overline{10}}$     & 1/2 & 1710 (input)\footnotemark[5]                  & $N(1710)P_{11}$\footnotemark[4] \\
$\Sigma_{\overline{10}}$& 1/2 & 1890\footnotemark[5]                          & $\Sigma(1660)P_{11}$\footnotemark[4] \\ 
$\Xi_{\overline{10}}$   & 1/2 & 2070\footnotemark[5]                          & $\Xi(1690)$(?) or $\Xi1950$(?)\footnotemark[4]\\
$\Theta_1$             & 3/2 & 1595                         & ---                 \\
$\Delta_{27}$     & 3/2 & 1750                         & $\Delta(1600)P_{33}$\footnotemark[4]\\
$N_{27}$          & 3/2 & 1746                         & $N(1720)P_{13}$ \footnotemark[4]    \\
$\Gamma_{27}$     & 3/2 & 1904                         & ---                 \\
$\Sigma_{27}$     & 3/2 & 1899                         & ---                 \\
$\Lambda_{27}$    & 3/2 & 1896                         & $\Lambda(1890)P_{03}$\footnotemark[4]\\
$\Pi_{27}$        & 3/2 & 2052                         & ---                  \\
$\Xi_{27}$        & 3/2 & 2048                         & $\Xi(1690)$(?) or $\Xi1950$(?)\footnotemark[4]\\
$\Omega_{27}$     & 3/2 & 2200                         & $\Omega(2250)$(?)\footnotemark[4]\\
$\Theta_2$             & 1/2 & 2028                         & ---   \\
$\Delta_{27}$     & 1/2 & 2246                         & $\Delta(1910)P_{31}$\footnotemark[4]\\
$N_{27}$          & 1/2 & 2189                         & ---   \\
$\Gamma_{27}$     & 1/2 & 2474                         & ---   \\
$\Sigma_{27}$     & 1/2 & 2411                         & ---   \\
$\Lambda_{27}$    & 1/2 & 2350                         & ---   \\
$\Pi_{27}$        & 1/2 & 2594                         & ---   \\
$\Xi_{27}$        & 1/2 & 2569                         & ---   \\
$\Omega_{27}$     & 1/2 & 2682                         & ---   \\
\end{tabular}
\end{ruledtabular}
\footnotetext[1]{From Ref.~\onlinecite{Nakano}.}
\footnotetext[2]{From Ref.~\onlinecite{Barmin}.}
\footnotetext[3]{From Ref.~\onlinecite{Stepanyan}.}
\footnotetext[4]{From Ref.~\onlinecite{PDG}.}
\footnotetext[5]{From Ref.~\onlinecite{DPP}.}
\end{table}

Because of the non-vanishing transition amplitudes (\ref{Transition}) we get second order corrections to the mass spectrum of the decuplet
\begin{equation}\label{2d_order}
\begin{split}
&\Delta m_\Delta^{(2)}=-\frac{15}{128}m_2,\\
&\Delta m_{\Sigma^\ast}^{(2)}=-\frac{1}{16}m_2,\\
&\Delta m_{\Xi^\ast}^{(2)}=-\frac{3}{128}m_2,\\
&\Delta m_\Omega=0, \quad m_2=I_1\left(\alpha+\frac56\beta\right)^2.
\end{split}
\end{equation}
These corrections lead to violations of the equidistance in the spectrum and to the following sum rules
\begin{equation}\label{New_relations}
\begin{split}
&m_{\Sigma^\ast}+m_{\Xi^\ast}-m_\Delta-m_\Omega=\\
&=2(2m_{\Xi^\ast}-m_{\Sigma^\ast}-m_\Omega)=\\
&=2(2m_{\Sigma^\ast}-m_\Delta-m_{\Xi^\ast}).
\end{split}
\end{equation} 
For an equidistant spectrum every line in (\ref{New_relations}) would give zero. But experimentally these quantities do not vanish. Inserting the experimental masses in these three lines results in $15~\mathrm{MeV/c^2}$, $16~\mathrm{MeV/c^2}$ and $13~\mathrm{MeV/c^2}$. This is in good agreement with the sum rules (\ref{New_relations}) derived from the second order corrections independent from the model parameters (\ref{Parameters}) and (\ref{First_MOI}). But inserting the values (\ref{Parameters}) and (\ref{First_MOI}) in the second order corrections (\ref{2d_order}) leads to deviations (\ref{New_relations}) from the equidistance two times smaller than in the experimental spectrum.

The second order corrections decrease the masses of the nucleon and $\Sigma$ in the octet by $5~\mathrm{MeV/c^2}$.

Using the baryon-meson coupling \cite{ANW}
\begin{equation}\label{Coupling} 
-i\frac{3G_0}{2m_B}\sum_{A=1}^3D^{(8)}_{mA}p_A,
\end{equation}
we have estimated the width of the $\Theta_1$-resonance in the non-relativistic limit. In (\ref{Coupling}) $\vec p$ is the meson momentum in the resonance frame and the coupling constant $G_0\approx 19$. The new family of $\Theta$-baryons, contrary to the $\Theta^+$, has normal hadronic width, $\Gamma\approx 80$~MeV. 

In conclusion, we predict that there exists a new isotriplet of $\Theta$-baryons, $\Theta_1^{++}$, $\Theta_1^+$ and $\Theta_1^0$, with hypercharge $Y=2$ and $J^P=\frac32^+$. Its mass and width, $1595~\mathrm{MeV/c^2}$ and 80~MeV, respectively, are predicted from the SU(3) Skyrme model using the same parameters as Diakonov et al. \cite{DPP} employed for the exotic $\Theta^+$ baryon which was recently observed experimentally \cite{Nakano,Barmin,Stepanyan}. The triplet of $\Theta_1$ baryons is a member of the 27-dimensional representation of the SU(3) flavor group. We identify other non-exotic members of this representation ($\Delta_{27}$, $N_{27}$, $\Lambda_{27}$ and, possibly, $\Xi_{27}$) with observed resonances, but do not see a structure, which can be related to the $\Sigma_{27}$ resonance. Further we predict two additional exotic resonances, $\Gamma_{27}$ and $\Pi_{27}$. It is shown that there exist strong mixtures between the decuplet and the 27-plet for states with quantum numbers of $\Delta$, $\Sigma$ and $\Xi$. These mixtures may be responsible for small violations of the equidistance in the decuplet spectra.

When the paper was finished there appeared an article by Jaffe and Wilczek \cite{JW}  where they propose that the $\Theta$-baryon ''lies in a near-ideally mixed $\mathrm{SU(3)_f\ \overline{10}_f\oplus 8_f} $''. The predicted spectrum differs essentially from the prediction of the Skyrme model. 

In another article, which appeared at the same time, $\Theta^+$ was discussed from the point of view of QCD sum rules \cite{Zhu}. A series of pentaquark states with isospin 0,1 and 2 with $J^P=\frac12^-$ is predicted to lie close to each other near 1550~MeV. In the Skyrme model we have also close resonances, $\Theta(1540)$ and $\Theta_1(1595)$, but with positive parity and spin $\frac12$ and $\frac32$, respectively, as given in  Table~\ref{tab:1}. 

The authors thank to Andro Kacharava and Eugene Strokovsky for helpful discussions.

\end{document}